\documentclass{pasj00}

\begin{document}
\SetRunningHead{T. Hagihara et al.}{X-Ray Spectroscopy of Galactic Hot Gas along the PKS 2155--304 Sight Line}
\Received{2009/12/16}
\Accepted{2010/3/29}
\Published{}

\title{X-ray Spectroscopy of Galactic Hot Gas along the PKS 2155-304 Sight Line}

\author{
Toshishige \textsc{Hagihara}\altaffilmark{1}, 
Yangsen \textsc{Yao}\altaffilmark{2}, 
Noriko Y. \textsc{Yamasaki}\altaffilmark{1}, 
Kazuhisa \textsc{Mitsuda}\altaffilmark{1},  \\
Q. Daniel \textsc{Wang}\altaffilmark{3},
Yoh \textsc{Takei}\altaffilmark{1},
Tomotaka \textsc{Yoshino}\altaffilmark{1}
\thanks{Present Address is NEC corporation, Nisshin-cho 1-10, Fuchu, Tokyo 183- 
8551} and 
Dan \textsc{McCammon}\altaffilmark{4} 
}
%
\altaffiltext{1}{Institute of Space and Astronautical Science, Japan Aerospace Exploration Agency, 3-1-1, Yoshinodai, Chuo, Sagamihara, 252-5210}
\email{hagihara@astro.isas.jaxa.jp, yamasaki@astro.isas.jaxa.jp}
\altaffiltext{2}{University of Colorado, CASA, 389 UCB, Boulder, CO 80309, USA}
\altaffiltext{3}{Department of Astronomy, University of Massachusetts,
  Amherst, MA 01003, USA}
\altaffiltext{4}{Department of Physics, University of Wisconsin, Madison, 1150 University Avenue, Madison, WI 53706, USA}
\email{hagihara@astro.isas.jaxa.jp, yamasaki@astro.isas.jaxa.jp}

%


\maketitle

\begin{abstract}
We present a detailed spectroscopic study of the hot gas in the Galactic halo 
toward the direction of a blazer PKS 2155-304 ($z=$0.117). 
The O\emissiontype{VII}  and O\emissiontype{VIII} absorption lines 
are measured with the Low and High 
Energy Transmission Grating Spectrographs aboard {\it Chandra}, 
and the O\emissiontype{VII}, O\emissiontype{VIII},  and Ne\emissiontype{IX}  emission lines 
produced in the adjacent field of the PKS 2155-304 direction are 
observed with the X-ray Imaging Spectrometer aboard {\it Suzaku}.
Assuming vertically exponential distributions of the gas temperature
and the density, we perform a combined analysis of the absorption and emission
data.  The  gas temperature and density at the
Galactic plane are determined to be
$2.5(+0.6,-0.3)\times 10^6$ K and $1.4(+0.5,-0.4)\times 10^{-3}$ cm$^{-3}$
and the scale heights of the gas temperature and density are 
$5.6(+7.4,-4.2)$ kpc and $2.3(+0.9,-0.8)$ kpc, respectively. 
These values are consistent with those obtained in the LMC X-3 direction.

\end{abstract}

\section{Introduction}
X-ray observations of edge-on spiral galaxies revealed the existence 
of hot gas at temperatures of $\sim$ 10$^{6}$ K extending a few kpc beyond the
disk (e.g. \cite{wang01, wang03, strickland04,li2008, yamasaki09}). 
The origin of energy and material in such a hot halo has not been clarified.
Feedback from supernovae (SNe)  as galactic wind or fountain and 
heated primordial gas  are possible candidates \citep{norman89}.
In any cases, halo gas plays important roles in
galactic evolution through chemical circulation and 
interaction between galaxies and the intergalactic medium.

The hot gaseous halo in and around the Milky-Way 
has been investigated for a long time. For instance, ROSAT All Sky Survey 
(RASS) quantitatively mapped the spatial distribution of 
the Soft X-ray Background emission (SXB; \cite{sno97}). 
The Cosmic X-ray Background (CXB) component extrapolated from 
the discrete hard X-ray sources 
could explain only about half of the SXB, leaving the soft X-ray emission
below 1 keV being of a diffuse origin. With the 
high resolution X-ray microcalorimeter flying on a sounding rocket,
\citet{mcc02} 
detected emission lines of hydrogen- and helium-like 
oxygen, neon, and iron ions from about 1 steradian  of the sky, which 
suggests that the emitting gas is of a thermal nature and at temperatures 
of T$\sim 10^6$ K. The existence of the hot gas in and around the Milky-Way
is consistent with the {\it Chandra} observations of nearby edge-on 
spiral galaxies. However, because these emission data carry very little 
distance information, the properties of the global hot gas, like its density,
temperature, and their distributions, are still poorly understood.

A combined analysis of high resolution absorption and emission data provides
us with a powerful diagnostic of properties of the absorbing/emitting plasma. 
Absorption lines measure the column density of the absorbing 
material, which is an integration of the density of the absorbing ions
along a sight line.
In contrast, emission line intensity is sensitive to the emission measure, 
which is proportional to the density square of the emitting plasma.
Thus, a combination of the emission and absorption data naturally yields
the density and the size of the corresponding absorbing/emitting gas.

With significantly improved spectral resolution of current X-ray 
instruments, we are now  able to observe the needed high resolution 
absorption and emission lines produced in the hot plasma. 
For instance, the X-ray absorption lines at $z=0$, in particular
the helium- and hydrogen-like O\emissiontype{VII} and O\emissiontype{VIII} lines, 
are detected in spectra of many galactic and 
extragalactic sources (e.g. \cite{fut04, yao05, wil07}).
Recently, \citet{fang06} and \citet{bre07} find that the O\emissiontype{VII}  absorption
line can always be detected in an AGN spectrum as long as the spectrum is
of high signal-to-noise ratio. On the other hand, the X-ray Imaging 
Spectrometer (XIS) aboard {\it Suzaku} can also resolve emission lines produced
in a diffuse emitting plasma at temperatures of T$\sim 10^6$ K. 
And indeed, the O\emissiontype{VII} and O\emissiontype{VIII} lines have been detected in nearly all 
directions (e.g., \cite{smith07, she07}). Recently, a systematical 
study of emission lines of the hot gas in and around the Galaxy has been 
conducted by \citet{yoshino09}, who report the O\emissiontype{VII} and O\emissiontype{VIII} lines
in 14 blank sky observations with the XIS and conclude that the line-of-sight mean 
temperatures of the  emitting gas has a narrow distribution around 
$2.3\times10^6$ K. 
Since the ion fractions of 
O\emissiontype{VII} and O\emissiontype{VIII} and their K-transition emissivities are very sensitive to 
gas temperature at 
$\sim10^6$ K 
, a combined
analysis of these emission and absorption lines will also constrain 
the gas temperature and its distribution without the complexity of 
relative chemical abundances of metal elements. 

Although this combined analysis method has long been applied in the ultraviolet 
wavelength band \citep{shu94}, its application in the X-ray band 
just began. Complementing the high resolution absorption data observed with
{\it Chandra} with the broadband emission data obtained with RASS, \citet{yao07}
firstly attempted to conduct the combined analysis in the X-ray band to infer
the hot gas properties in our Galaxy. They also proposed a model for the 
Galactic disk assuming the temperature and density of the hot gas fading off
exponentially along the vertical direction. They concluded that the 
O\emissiontype{VII}  and O\emissiontype{VIII}  absorption lines observed along the Mrk~421 sight line
are consistent with the Galactic disk origin.  
\cite{yao09} further constrained this disk model by
jointly analyzing the high resolution absorption data obtained
with {\it Chandra} along the LMC~X-3 sight line and emission data 
observed with {\it Suzaku} in the vicinity of the sight line. They
estimated gas temperature and density at
the Galactic plane and their scale heights as 3.6 (+0.8, $-$0.7) $\times 10^6$
K and 1.4 (+2.0, $-$1.0) $\times 10^{-3}$ cm$^{-3}$ and 1.4 (+3.8, $-$1.2) kpc and
2.8 (+3.6, $-$1.8) kpc, respectively.  These results are consistent with the
early findings by \citet{yao07}, i.e., the SXB can be 
explained by a kpc-scale halo around our Galaxy. 

In this paper, we present the second case study of the combined analysis 
of high resolution absorption and emission lines.
The absorption lines are observed with {\it Chandra}  along a blazer, 
PKS~2155--304 sight line and the emission lines are obtained with {\it Suzaku}
observations of the vicinity of the sight line. 
In Section 2, we describe our observations and data reduction process.
We perform our data analysis in Section 3 and discuss 
our results in Section 4.

\section{Observations and Data Reduction}

\begin{table*}
\begin{center}
 \caption{{\it Suzaku} Observation Log}
 \label{tb:SuzakuObsLog}
\begin{tabular}{lcc}\hline \hline
  & Sz1 & Sz2 \\ \hline
($\alpha$, $\delta$) in J2000 (degrees) & (329.2236,$-$30.5193) & (330.1731,$-$29.9560)\\
($\ell$,$b$) in Galactic coordinate (degrees) & (17.1809,$-$51.8544) & (18.2418,$-$52.6081)\\
 Observation ID & 503082010 & 503083010 \\
 Observation start times (UT)  & 18:32:39, 2008 Apr 29 & 08:31:41, 2008 May 2\\
 Observation end times (UT) & 08:30:08, 2008 May 2 & 17:30:19, 2008 May 4\\
 Exposure time  & 90ks & 87ks \\
 Exposure after the data reduction & 51.1ks & 56.3ks \\\hline
\end{tabular}
\end{center}
\end{table*}

\subsection{{\it Chandra} Observations and Data Reduction}
{\it Chandra}  observed PKS~2155--304 many times. There are two grating 
spectrographs (the low and high energy transmission grating spectrographs;
LETG and HETG) and two sets of detectors (
the advanced CCD imaging spectrometer; ACIS and the high resolution
camera; HRC) 
aboard {\it Chandra}
\footnote{please refer to the Chandra Observatory Guide for more
information:\\ http://cxc.harvard.edu/proposer/POG/html/index.html}.
In this work, we used all observations available to the date of
2009 March, except for some observations made with 
non-standard configuration of ACIS (i.e., putting source outside the CCD-S3 
chip) to avoid spectral resolution degradation. The data
used in this work include 46 observations with an accumulated 
exposure time of 1.07 Ms.

We followed the standard scripts to calibrate the observations
\footnote{Please refer to the CIAO script for more
information: http://cxc.harvard.edu/ciao/guides/}. When extracting 
the grating spectra and calculating the instrumental response files, 
we used the same energy grid for all observations
with different grating instruments and/or with different detectors
for ease of the adding process described in the following.
For those HETG observations, we only use the first order grating 
spectra of the medium 
energy grating (MEG) to utilize its large effective area at lower ($<1$ keV) energy.
For those observations taken with the HRC, we further followed the procedure
presented in \citet{yao09} to extract the first order spectra of
the LETG. We then added the first grating order 
spectra of all observations to obtain a 
single stacked spectrum and a corresponding instrumental response file.

\subsection{Suzaku Observations and Data Reduction}

\begin{figure}
\begin{center}
\end{center}
\FigureFile(80mm, 50mm){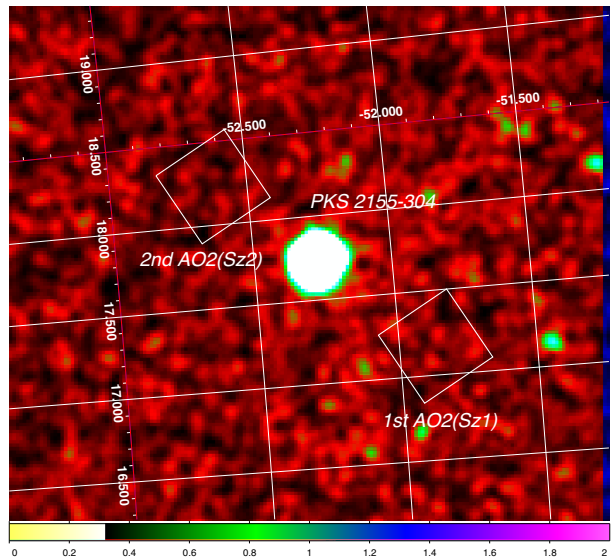}
\caption{RASS 0.1 - 2.4 keV band X-ray map in the vicinity
 of PKS 2155-304 (the bright source at the center) and the XIS
 field of view of the two presented observations. }
\label{fig:SuzakuFOV} 
 \end{figure}

\begin{figure}[h]
\FigureFile(80mm,50mm){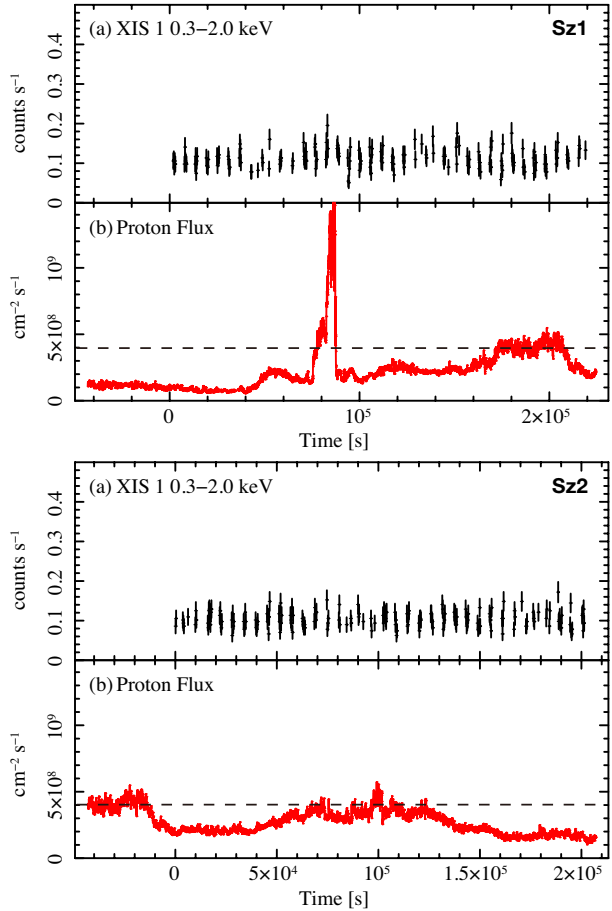}
\caption{ (a) XIS light curve in 0.3-2.0 keV  and (b) solar wind
 proton flux  calculated using the data of ACE SWEPAM 
in Sz1 (top) and Sz2 (bottom) observation
 periods. The time is plotted from the beginning of each observation with
 {\it Suzaku}. The time bin of proton data are shifted 5000 seconds to
 correct for the travel time of the solar wind from the ACE satellite to
 the Earth.
The dashed lines in the bottom panel indicate the
threshold of the proton flux as $4 \times 10^8$  cm$^{-2}$ s$^{-1}$.
}
\label{fig:LightCurve} 
\end{figure}

We observed the emission of the hot diffuse gas toward two 
off-fields of the PKS 2155-304 sight line during the AO2 program 
(Table \ref{tb:SuzakuObsLog}). To minimize confusion by stray lights from
the PKS 2155-304 and to average out the possible spatial gradient of the 
diffuse emission intensity, the two fields were chosen to be
30$'$ away from the PKS~2155--304 and in nearly opposite directions 
(Fig. \ref{fig:SuzakuFOV}).
With this configuration and the roll 
angle of the XIS field of view, we estimate that stray lights from 
PKS 2155-304 contribute no more than 10\% to the observed X-ray 
emission in 0.3--1.0 keV energy range.
Our observation pointings are away from the southern edge of 
Radio Loop I \citep{Berkhuijsen}.   Thus we consider that there is no
contamination of the emission from the Loop I in our observations.  This
is supported by the observational results that there is no EUV
enhancement in this direction \citep{Sembach97}.

Our observations were taken with the CCD camera 
X-ray Imaging Spectrometer (XIS;
\cite{koyama07}) on board Suzaku \citep{mitsuda07}. 
The XIS was set to the normal clocking mode and the data 
format was either $3\times3$ or $5\times5$, and the spaced-raw charge 
injection (SCI) was applied to the data during the observations. 
We used processed data version 2.2.7.18 for the two observations.
In this work, we only used the spectra obtained with XIS1.
Compared to the other two front side-illuminated
CCDs, XIS0 and XIS3, XIS1 is a
backside-illuminated CCD chip and is of high sensitivity at photon
energies below 1 keV. We found no point sources in the FOV, thus we used the full CCD field of view in further
analysis to increase the photon counts
because X-ray from the calibration sources do not
affect the soft X-ray spectrum below 5 keV.

We adopted the standard data selection criteria to obtain 
the good time intervals (GTIs), i.e. excluding exposures when the
line of sight of {\it Suzaku} is elevated above the Earth rim by less than
20$^\circ$
and exposures with the ``cut-off rigidity'' less than 8 GV.
We checked the column density of the neutral oxygen in the Sun-lit
atmosphere in the line of sight during the
selected GTIs, and excluded the exposures when the column density is
larger than $1.0 \times 10^{15}$ cm$^{-2}$  
to avoid 
significant neutral oxygen emission from Earth's atmosphere 
\citep{smith07}.
We created X-ray images in 0.4--1.0 keV energy range for the two
observations, and found no obvious discrete X-ray sources in the fields.

In the last step, we excluded those events 
severely contaminated by the X-ray emission
induced by the solar-wind charge exchange (SWCX)
from geocorona (Fujimoto et al. 2007), meeting either of the following 
two criteria by \citet{yoshino09}.  The first one is the solar wind flux
(Fig.\ref{fig:LightCurve}).
We used the solar wind data obtained with the Solar
Wind Electron Proton and Alpha Monitor (SWEPAM) aboard the 
{\it Advanced Composition Explorer} ({\it ACE }) and removed the time intervals
when the proton flux in the solar wind exceeds 
$4 \times 10^8$  cm$^{-2}$ s$^{-1}$  \citep{masui09}.
ACE is in L1 of the Solar-Earth system, 1.5$\times 10^6$ km away from the
Earth and assuming average solar wind velocity as 300 km s$^{-1}$, we
corrected traveling time of the solar wind from L1 to the Earth.
The second criteria is the Earth-to-magnetopause (ETM) distance in
the line sight of Suzaku \citep{fujimoto07}, which is required to
be $> 5 R_E$. 
We found that about 20\% and 5\% of the exposure time of our 
1st and 2nd observations meets the first criteria and no time meets the
second criteria.
Thus we exclude that 20\% and 5\% from 1st and 2nd observations and
used the remaining time in further analysis.
We also checked the light curve of XIS 1 in the energy range of 0.3 to
2.0 keV in the observation periods and found no evidence 
for variation (Fig. \ref{fig:LightCurve}).

We constructed instrumental response files (rmfs) and effective area
files  (arf) by running the scripts {\it xisrmfgen} and {\it
xissimarfgen} \citep{ishisaki07}.
 To take into account the diffuse stray light effects, we
used a 20$''$ radius flat field as the input emission in calculating the
arf. We also included in the arf file the degradation of low energy
efficiency due to the contamination on the XIS optical blocking filter.
The versions of calibration files used here were
ae\_xi1\_quanteff\_20080504.fits, ae\_xi1\_rmfparam\_20080901.fits,
ae\_xi1\_makepi\_20080825.fits and ae\_xi1\_contami\_20071224.fits.
We estimated the non-X-ray-background from the night Earth database
using the method described in \citet{tawa08}.

 We grouped the spectra to have a 
minimum number of counts in each channel $\geq$ 50  and used energy range of
0.4--5.0 keV in
our analysis. This range is broad enough for constraining the continuum
and also covers the H- and He-like emission lines of N,
O, Ne, Mg, and the L transition of Fe.
The O\emissiontype{VII}, O\emissiontype{VIII},  and Ne\emissiontype{IX}  lines are clearly visible in the
spectra (Section 3.2).

\section{Spectral Analysis and Results}
We carried out our data analysis with the Xspec software package, adopting
the solar abundances as given in  \citet{and89}. 
(Hereafter, use of {\it italic} type indicates Xspec models and their parameters.) 
Errors quoted throughout this paper are single parameter errors
given at the 90 \% confidence level, unless specified otherwise.
Sections 3.1 and 3.2 give a discussion of our separate analyses of
the absorption and emission data, while sections 3.3 and 3.4 give a 
discussion of the jointly--analyzed data under the uniform and exponential
disk models.

\subsection{ {\it Chandra} X-ray Absorption Spectrum} 

We first measured the equivalent widths (EWs) of the absorption lines
of the highly ionized oxygen ions. Becuase measurement of these 
narrow absorption lines is relevant only to the local continuum, 
we fit the final PKS 2155-304 spectrum between 0.55 and 0.7 keV as shown in figure 3
using a power-law  model modified with absorption by the neutral ISM({\it wabs}).
The column density of neutral hydrogen was fixed to  
1.47$\times 10^{20}$~cm$^{-2}$, which is the value determined 
by the LAB Survey of Galactic HI in this direction \citep{Kalberla}. 
Three Gaussian functions were used to model the O\emissiontype{VII} K$_{\alpha}$, 
 O\emissiontype{VIII} K$_{\alpha}$, and  O\emissiontype{VII} K$_{\beta}$
 absorption lines (model A1). The measured EWs were found to be consistent
with those reported by \cite{wil07}. The results are 
summarized in Table~\ref{tb:FitResultsAbs1}

Once the equivalent widths were determined, we applied an absorption line
model, {\sl absem}, to replace the {\sl gaussian} functions in order to probe 
the properties of the absorbing gas. Assuming the temperature and density distributions
of the hot plasma, the {\sl absem} model, which is a revision of the 
{\sl absline} model of \citet{yao05}, can be used to jointly fit the emission and absorption spectra.
(See \citet{yao07} and \citet{yao09} for a detailed description.) For a gas with a uniform density
and a single temperature,  the diagnostic procedure is summarized as follows: (1)
 A joint analysis of O\emissiontype{VII} K$_{\alpha} $ and 
O\emissiontype{VII} K$_{\beta}$ directly constrains
the O\emissiontype{VII} column density and the Doppler dispersion velocity
($v_{\rm b}$). With the constrained $v_{\rm b}$, adding the
O\emissiontype{VIII} K$_{\alpha}$ line
in the analysis also yields the column density of O\emissiontype{VIII} (model A2).
(2) Because the column density ratio of O\emissiontype{VII} and O\emissiontype{VIII} 
is sensitive  to the gas temperature, a joint analysis of the O\emissiontype{VII} and 
O\emissiontype{VIII} lines will naturally constrain the gas temperature (model A3).
(3) Assuming the solar abundance for oxygen and given the constrained gas temperature,
the O\emissiontype{VII} (or O\emissiontype{VIII} ) column density can be converted to the corresponding
hot phase hydrogen column density (model A4). Table~\ref{tb:FitResultsAbs2} gives the results of our fits.
The constrained O\emissiontype{VII} column density, 
5.9 (+1.2, $-$0.9) $\times 10^{15}$~cm$^{-2}$ is comparable to 
typical values $\sim10^{16}$ cm$^{-2}$ obtained from AGN observations given in two 
systematic studies (\citet{fang06} and \citet{bre07}).

\begin{figure}[h]
\begin{center}
\FigureFile(80mm, 50mm){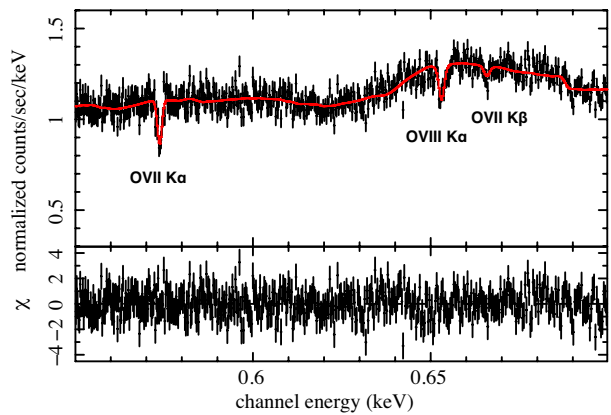}
\end{center}
\caption{Chandra spectrum of PKS 2155-304 between 0.55 and 0.7
 keV. Fitted model is A4.}
\label{fig:absorption} 

\end{figure}

\begin{table*}[htb!]
 \begin{center} 
 \caption{Spectral fitting results of absorption data with model A1}
 \label{tb:FitResultsAbs1}
\begin{tabular}{llccc}\hline \hline
Model& & O\emissiontype{VII} K$_{\alpha}$ &O\emissiontype{VIII} K$_{\alpha}$   &O\emissiontype{VII} K$_{\beta}$   \\ \hline
A1	& Centroid (eV) & $573.8^{+0.1}_{-0.2}$ & $653.1^{+0.4}_{-0.4}$ &
 $665.8^{+0.1}_{-0.4}$ \\ 
 & Sigma (eV) & $0.32^{+0.25}_{-0.32}$ & $1.01^{+0.62}_{-0.54}$ &
 $0.01^{+0.97}_{-0.01}$ \\ 
  & Equivalent Width (eV) & $0.354^{+0.075}_{-0.071}$  &  $0.377^{+0.116}_{-0.102}$ & $0.119^{+0.058}_{-0.058}$	 \\ \hline
\multicolumn{4}{l}{{\footnotesize Model A1:{\it wabs}({\it power-law}+$3\times${\it Gaussian})}}
\end{tabular}
 \end{center}
\end{table*}

\begin{table*}[htb]
\begin{center}  
 \caption{Spectral fitting results of absorption data with model A2-A4}
 \label{tb:FitResultsAbs2}
\begin{tabular}{lcccccc}\hline \hline
 Model &  $v_{b}$  & \multicolumn{3}{c}{$\log$ [Column Density] }  & $\log T$& $\chi^2$/dof\\ 
&  (km s$^{-1}$) & \multicolumn{3}{c}{  (cm$^{-2}$) }  &  (K) & \\ 
 & &  $N_{\rm O\emissiontype{VII}}$ & $N_{\rm O\emissiontype{VIII}}$  & $N_{\rm H_{Hot}}$  & &\\ \hline
 A2 &  $294^{+149}_{-220}$ & $15.76^{+0.07}_{-0.08}$ &
				 $15.56^{+0.09}_{-0.12}$ & $\cdots$ & $\cdots$ &489.82/474\\ 
 A3  & $375^{+124}_{-158}$ & $15.77^{+0.08}_{-0.07}$
			 &$\cdots$ & $\cdots$ & $6.27^{+0.02}_{-0.03}$ & 498.01/474\\
 A4 & $290^{+152}_{-220}$ & $\cdots$ & $\cdots$ &
			$19.08^{+0.06}_{-0.07}$ & $6.28^{+0.02}_{-0.02}$ &
 489.84/474\\ \hline
 \multicolumn{6}{l}{ {\footnotesize Model A2,A3,A4:{\it wabs}({\it power-law})$\times${\it absem}$\times${\it absem}$\times${\it absem}}}
\end{tabular}
\end{center}
\end{table*}

\subsection{ {\it Suzaku} X-ray Emission Spectra }

The {\it Suzaku} data were modeled in order to constrain 
the emission measure and the temperature of the halo.
For this purpose, we first modeled the SXB using a multiple component
model, since the SXB emission is a superposition of such components. 
We detail this model below.

\subsubsection{Foreground and Background Emission}
We assumed that the SXB consists of four dominant components:
(1) the Local Hot Bubble (LHB),  (2) Solar Wind Charge eXchange 
in the heliosphere (SWCX),  (3) a hot gaseous Galactic halo,  
(4) the cosmic X-ray background emission (CXB; mainly from unresolved
extragalactic sources such as AGNs).
Because the contribution from unresolved Galactic sources is expected to be
negligible at high galactic latitudes ($|b|>30^{\circ}$), 
we did not consider such a contribution.
The CXB spectrum is well described by a {\it power-law}. 

In a study of 14 {\it Suzaku} blank sky observations, 
\citet{yoshino09} found that there are at least 2 LU 
(photons ${\rm s^{-1} cm^{-2} str^{-1}}$) of OVII line emission 
even in those directions where the
attenuation length for the line is less than 300 pc. This emission is
considered to come from the SWCX and LHB, though these contributions are difficult
 to separate with the current CCD energy resolution. After \citet{smith07}
and \citet{hen07},
\citet{yoshino09} found that 
it can be well represented by a model consisting of unabsorbed, optically 
thin thermal emission from a collisionally-ionized plasma. The best-fit
temperature of this model is log $T$ = 6.06. 
We therefore use a $\sim10^6$ K plasma of 2 LU O\emissiontype{VII} surface
brightness as the SWCX+LHB component. 
The uncertainty of this estimate is discussed in section \ref{sec:LHBandSWCX}.

Except for the SWCX+LHB component, the observed 
emission has been absorbed by the foreground ISM.
In the following analysis, we also fix
the neutral hydrogen column density to be 
$1.47\times10^{20}~{\rm cm^{-2}}$ (\cite{Kalberla}).

\begin{figure}[h]
\FigureFile(80mm,50mm){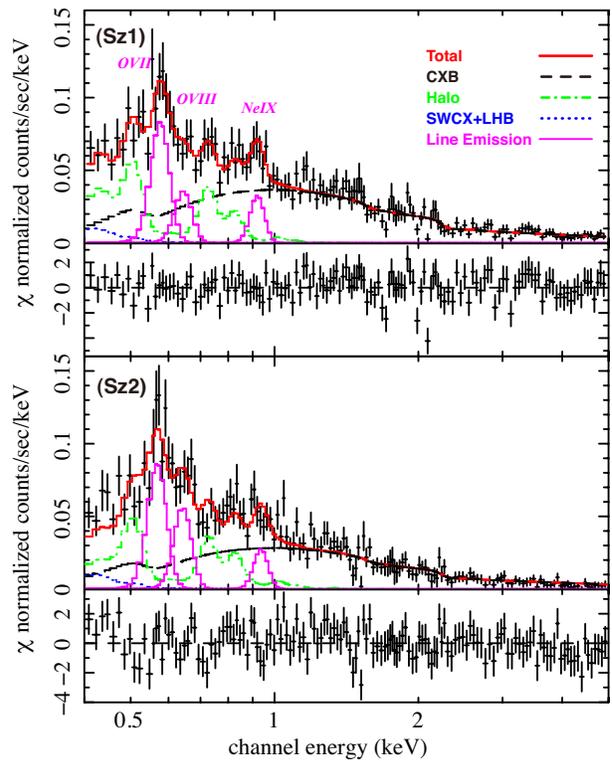}
\caption{Suzaku spectra between 0.4 and 5.0 keV of Sz1 (top) and Sz2 (bottom)
  are  plotted. 
Fitted model is E2 ({\it wabs}({\it power-law}$_{CXB}$ + {\it
 vmekal}$_{halo}$) + {\it vmekal}$_{LHB+SWCX}$ + 3$\times${\it
 gaussians}).
The O and Ne abundance of the {\it
 vmekal}$_{halo}$ (green, dash-dotted) and {\it vmekal}$_{LHB+SWCX}$
 (blue, dotted) are set to be {\sl zero}
 and three {\it gaussians} (magenta, solid)
 represent O\emissiontype{VII} K$_{\alpha}$,
(O\emissiontype{VII} K$_{\beta}$ + O\emissiontype{VIII} K$_{\alpha}$) and
Ne\emissiontype{IX} K$_{\alpha}$ emission lines.
}
\label{fig:emissionGa} 
\end{figure}

\begin{figure}[h!]
\FigureFile(80mm,50mm){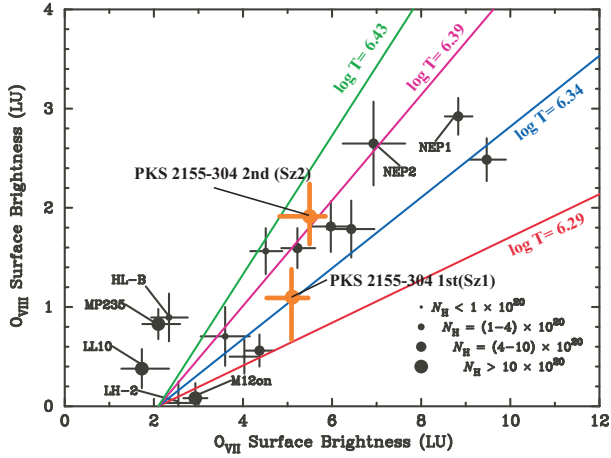}
\caption{Relation between O\emissiontype{VII} and O\emissiontype{VIII}  surface brightnesses for the
 14 (\cite{yoshino09})+2 (this work) sky fields observed with {\it Suzaku}. 
The horizontal and vertical bars of data points show the 1 $\sigma$
 errors of the estimate. The contribution 
of O\emissiontype{VII} K$_{\beta}$ emission is corrected for
 O\emissiontype{VIII} K$_{\alpha}$. The 
diagonal lines show the relation between O\emissiontype{VII}and O\emissiontype{VIII}, assuming an offset O\emissiontype{VII} emission of 2.1 LU and 
emission from a hot plasma of the temperature and the absorption column
 density are shown. 
The Galactic absorption column density of the observation fields are indicated by the maker size of the 
data points. 
}
\label{fig:OviiVsOviii} 
\end{figure}

\begin{table*}
\begin{center}  
 \caption{Spectral fitting results for emission data with the model E1}
 \label{tb:FitResultsEmi2}
\begin{tabular}{llcccccccccc}\hline \hline
Model & Data & CXB &
 \multicolumn{2}{c}{LHB+SWCX} & \multicolumn{5}{c}{halo} & $\chi^2$/dof\\ 
 &   & Norm $^{a}$ & $\log T$  & Norm$^{b}$ & $\log T$ & Norm$^{b}$ & N/O & Ne/O & Fe/O \\ 
&   &  & (K) &  & (K) &  & & & \\ \hline
 E1 & Sz1  & $8.40^{+0.38}_{-0.40}$  & 6.06(fixed) & 4.3(fixed) &
					 $6.26^{+0.06}_{-0.04}$ & $3.3^{+1.0}_{-0.8}$ &
 $6.0^{+2.5}_{-1.9}$ & $6.5^{+3.7}_{-2.5}$ & $7.4^{+13.8}_{-4.8}$ &
 148.61/136\\
 E1 & Sz2  & $6.45^{+0.36}_{-0.43}$ & 6.06(fixed) & 4.3(fixed) &
					 $6.35^{+0.03}_{-0.03}$ & $3.2^{+0.5}_{-0.4}$ & 
 $4.7^{+2.0}_{-1.6}$ & $2.4^{+1.2}_{-0.9}$ & $1.0^{+0.8}_{-0.5}$ &
 147.33/141\\
 E1 & Sz1+Sz2(Sz1) & $8.30^{+0.35}_{-0.39}$ & 6.06(fixed) & 4.3(fixed) &
					 $6.33^{+0.02}_{-0.02}$ & $3.0^{+0.3}_{-0.3}$ & 
 $5.8^{+1.6}_{-1.3}$ & $3.3^{+1.2}_{-0.9}$ & $1.7^{+1.2}_{-0.7}$ & 306.82/282\\ 
 & Sz1+Sz2(Sz2) & $6.50^{+0.36}_{-0.39}$ & $\uparrow$ & $\uparrow$ & 
					  $\uparrow$ & $\uparrow$  & 
  $\uparrow$ & $\uparrow$ & $\uparrow$ &  \\ \hline
 E1 $^{\dagger}$ & Sz1+Sz2(Sz1) & $8.38^{+0.35}_{-0.36}$ & 6.06(fixed) & 0.0(fixed) &
					 $6.25^{+0.03}_{-0.02}$ & $4.9^{+0.7}_{-0.6}$ & 
 $4.2^{+1.0}_{-0.8}$ & $4.5^{+1.4}_{-1.2}$ & $4.7^{+3.0}_{-1.6}$ & 313.48/282\\ 
 & Sz1+Sz2(Sz2) & $6.59^{+0.34}_{-0.39}$ & $\uparrow$ & $\uparrow$ & 
					  $\uparrow$ & $\uparrow$  & 
  $\uparrow$ & $\uparrow$ & $\uparrow$ &  \\ \hline
 E1 $^{\ddagger}$ & Sz1+Sz2(Sz1) & $8.25^{+0.38}_{-0.37}$ & 6.06(fixed) & 7.5(fixed) &
					 $6.37^{+0.03}_{-0.03}$ & $2.3^{+0.3}_{-0.3}$ & 
 $6.9^{+2.4}_{-1.9}$ & $3.2^{+1.3}_{-1.0}$ & $1.5^{+0.9}_{-0.5}$ & 299.45/282\\ 
 & Sz1+Sz2(Sz2) & $6.47^{+0.36}_{-0.38}$ & $\uparrow$ & $\uparrow$ & 
					  $\uparrow$ & $\uparrow$  & 
  $\uparrow$ & $\uparrow$ & $\uparrow$ &  \\ \hline
\multicolumn{11}{l}{{\footnotesize $\uparrow$ indicates linked parameters}} \\
\multicolumn{11}{l}{{\footnotesize Sz1+Sz2: simultaneous fitting of the data Sz1 and Sz2}}\\
\multicolumn{11}{l}{ {\footnotesize Model E1:{\it wabs}({\it power-law}$_{CXB}$ + {\it vmekal}$_{halo}$) + {\it mekal}$_{LHB+SWCX}$}} \\
\multicolumn{11}{l}{{\footnotesize Emission measure of {\it mekal} $_{LHB+SWCX}$ is fixed to 0.0043cm$^{-6}$
 which corresponds to 2.0 LU of O\emissiontype{VII} K$_{\alpha}$ emission}} \\
\multicolumn{11}{l}{$^{\dagger}$ {\footnotesize Emission measure of {\it mekal}$_{LHB+SWCX}$ is set to 0 as the lower limit }} \\
\multicolumn{11}{l}{$^{\ddagger}$ {\footnotesize Emission measure of {\it mekal}$_{LHB+SWCX}$ is set to the upper limit which corresponds to 3.5 LU of  O\emissiontype{VII} K$_{\alpha}$ emission}} \\
\multicolumn{11}{l}{$^{a}$ {\footnotesize in unit of photons cm$^{-2}$ s$^{-1}$str$^{-1}$ eV$^{-1}$ @1keV}} \\
\multicolumn{11}{l}{$^{b}$ {\footnotesize Emission measure 10$^{-3}$ $\int n_e n_p dl$: in unit of cm$^{-6}$ pc}} \\
\end{tabular} 
\end{center}
\end{table*}

\begin{table*}
\begin{center}  
 \caption{Surface brightness of O\emissiontype{VII}, O\emissiontype{VIII} and Ne\emissiontype{IX}}
 \label{tb:SurfaceBrightness}

\begin{tabular}{llcccccccccc}\hline \hline
Model & Data & CXB 
 & \multicolumn{3}{c}{halo} &\multicolumn{1}{c}{O\emissiontype{VII} K$_{\alpha}$$^{c}$} &
 \multicolumn{1}{c}{O\emissiontype{VII} K$_{\beta}$+} & \multicolumn{1}{c}{O\emissiontype{VIII} K$_{\alpha}$$^{c}$}&
 Ne\emissiontype{IX} K$_{\alpha}$$^{c}$ &$\chi^2$/dof\\ 
&  & Norm $^{a}$ &  Norm$^{b}$ & N & Fe && O\emissiontype{VIII} K$_{\alpha}$$^{c}$&  &  \\ \hline
 E2 & Sz1  & $8.21^{+0.62}_{-0.27}$
					 & $4.2^{+0.3}_{-0.8}$
 &$6.0$ (fixed) &7.4 (fixed)
						 &$5.00^{+0.69}_{-0.80}$
							 &$1.45^{+0.33}_{-0.51}$
								 &$1.10^{+0.39}_{-0.56}$
								 &$0.65^{+0.12}_{-0.26}$
 &136.39/132\\
 E2 & Sz2  & $6.37^{+0.53}_{-0.26}$
		 & $4.5^{+0.7}_{-0.6}$
				 &  4.7 (fixed) &1.0 (fixed)
						 &$5.15^{+0.66}_{-0.86}$ 
							 &$1.98^{+0.53}_{-0.37}$ 
								 &$1.62^{+0.59}_{-0.42}$
								 &$0.58^{+0.10}_{-0.29}$
&150.59/137\\
\hline
\multicolumn{11}{l}{{\footnotesize model E2: {\it wabs}({\it power-law}$_{CXB}$ + {\it vmekal}$_{halo}$) + {\it vmekal}$_{LHB+SWCX}$ + 3$\times${\it gaussians}, where
O and Ne abundances of two {\it vmekal} are set to 0}} \\
\multicolumn{11}{l}{$^{a}$ {\footnotesize in unit of photons cm$^{-2}$ s$^{-1}$str$^{-1}$ eV$^{-1}$ @1keV}} \\
\multicolumn{11}{l}{$^{b}$ {\footnotesize Emission Measure 10$^{-3}$ $\int n_e n_p dl$: in unit of cm$^{-6}$ pc}} \\
\multicolumn{11}{l}{$^{c}$ {\footnotesize in unit of LU = photons
s$^{-1}$ cm$^{-2}$ str$^{-1}$}} \\\end{tabular} 
\end{center}
\end{table*}

\subsubsection{Spectral Fitting}
To probe the halo gas properties, we used the following model to fit
our spectra (model E1): \\
${\it wabs(power-law _{CXB} + vmekal_{halo}) + mekal_{LHB+SWCX}}$,
with the photon index of the CXB fixed at 1.4 and with the normalization as a free
parameter. The temperature and the corresponding emission measure (and thus the
normalization) of the {\it mekal}$_{LHB+SWCX}$ component
were set to $1.2 \times 10^6$ K and $0.0043$ pc cm$^{-6}$, respectively,
correspondind to 2 LU of  O\emissiontype{VII} K$_{\alpha}$ line emission.  
In the halo component, we fixed the abundance ratio of oxygen to hydrogen
to the solar value, and allowed the abundances of 
nitrogen, neon, and iron vary. 

This model fit the spectra from both pointings 
consistently, except for an
apparently higher neon and iron abundance in Sz1
(Table~\ref{tb:FitResultsEmi2}) which
 would be caused by a lower temperature 
of the Sz1 halo component. It is important 
to clarify whether this is caused by statistical
effects or by a true difference in plasma temperature.
The surface brightness of each line is a better indicator for this purpose.

We next evaluated the surface brightness of O\emissiontype{VII} and O\emissiontype{VIII} lines 
by modifying model E1 (this is model E2).
We set the O and Ne abundance  of the halo and LHB+SWCX  to {\sl zero}
 and used three {\it Gaussian}
emission lines to represent O\emissiontype{VII} K$_{\alpha}$,
(O\emissiontype{VII} K$_{\beta}$ + O\emissiontype{VIII} K$_{\alpha}$) and
Ne\emissiontype{IX} K$_{\alpha}$ emission
(Fig. \ref{fig:emissionGa}). 
Since the XIS resolution is not high enough to enable us to distinguish the OVII K$_{\beta}$ (656 eV) 
and OVIII K$_{\alpha}$ (653 eV) lines, they were modeled as a single line.
This model fitted  both spectra with a $\chi^2$/dof of 135.52/132 and
150.59/137 respectively.
Assuming the ratio between O\emissiontype{VII} K$_{\beta}$, and
O\emissiontype{VII} K$_{\alpha} (=\mu)$ intensities is
0.07 (see footnote 3) \footnotetext[3]{$\mu$ is a slow function of the plasma temperature for
thermal emission and here the value is 0.056. If the emission is due to
SWCX, $\mu=0.083$ \citep{2003ApJ...585L..73K}. We averaged these two values and used $\mu=0.07$
here. See \citet{yoshino09} section 3.1 for details.},
we calculated the O\emissiontype{VII}, O\emissiontype{VIII} and Ne\emissiontype{IX}  surface brightnesses
 as listed in Table~\ref{tb:SurfaceBrightness}. 
Intensities of these lines between the two fields are consistent to within the 90\%
confidence level, 
and we assume that the temperature difference is not essential.
We plotted  the O\emissiontype{VII} and O\emissiontype{VIII}  surface brightness over the 
\citet{yoshino09} results (Fig. \ref{fig:OviiVsOviii}, with 1 $\sigma$ error) and found that the O\emissiontype{VII} and 
O\emissiontype{VIII} surface brightness
of the PKS 2155-304 direction matches the trend of the other 14 fields.

\setcounter{footnote}{3}

\begin{figure}[h!]
\FigureFile(80mm,50mm){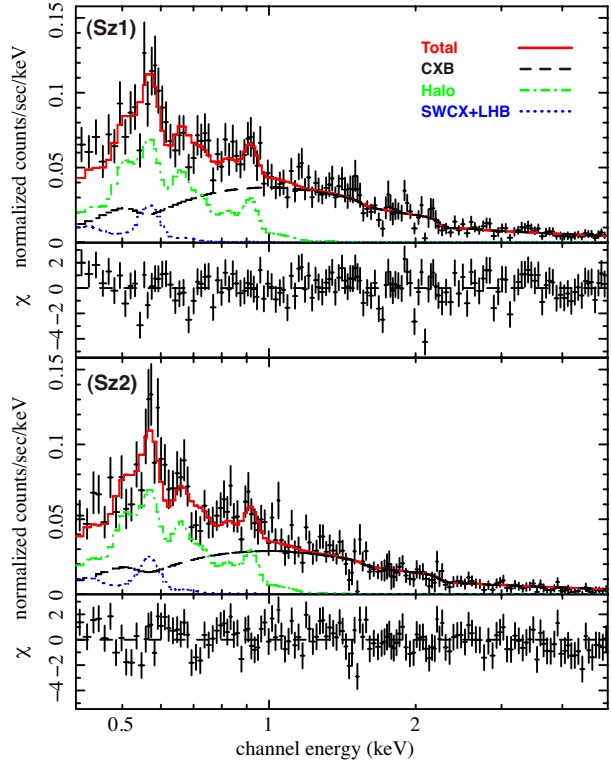}

\caption{Suzaku spectra between 0.4 and 2.0 keV. Sz1 (top) and Sz2 (bottom)
 observations are plotted. 
Fitted model is E1 ({\it wabs}({\it power-law}+{\it vmekal} $_{halo}$) +{\it mekal}$_{LHB+SWCX}$) and
parameters of the halo components are linked in both spectra.
}
\label{fig:emission} 
\end{figure}

We next
fitted both data sets simultaneously with model E1 by
linking parameters of the halo component in both observations.
The results are shown in Table~5.
(Fig. \ref{fig:emission}). 
The emission measure for the model is 3.0 (+0.3, $-$0.3) $\times 10^{-3}$ cm$^{-6}$ pc and
the temperature is 2.1 (+0.1, $-$0.1) $\times 10^{6}$ K.
\citet{mcc02} reported the emission measure and temperature of the absorbed
 thermal component (=halo) as  
3.7 $\times 10^{-3}$ cm$^{-6}$ pc and  2.6  $\times 10^{6}$ K which are
comparable to our values.

\subsection{Combined Analysis}
Up to now, we have analyzed the absorption and emission data separately and 
confirmed that the models including the halo component fit both
data with a temperature of 1.91(+0.09, $-$0.09) $\times 10^6$ K for the absorption
and 2.14 (+0.15, $-$0.14) $\times 10^6$ K for the emission spectra.

Assuming that both plasmas are common and uniform,  the plasma length and density can be calculated 
using the emission measure and the column density. The length and density are found to 
be 4.0 (+1.9, $-$1.4) kpc and 7.7 (+2.3, $-$1.7) $\times 10^{-4}$
cm$^{-3}$, respectively.
The errors of the calculated values are overestimated, since these errors
are not independent.
Moreover, important plasma parameters such as temperature and velocity
dispersion were not considered in this simple calculation.

In this section, using the combined analysis, we will try to determine the
physical conditions of the halo plasma, including the density, the temperature and
their distribution.

\subsubsection{Uniform Disk Model}
The first step in our combined analysis was to try the simplest model:
an isothermal plasma with uniform density extending up to $h$ kpc above
the disk (model C1).

To perform this combined analysis the emission measure and column density 
have to be linked with a common parameter. We chose the equivalent
hydrogen column
density ($N_{\rm H_{Hot}}$) and scale height ($h$) as the control
parameters and calculated the emission measure. The relation of the
density $n$, scale height $h$, column
density $N_{\rm H_{Hot}}$ and galactic latitude $b$ is described as
$N_{\rm H_{Hot}}=n h/{\rm sin}b$.
Thus we can use the A4 model for absorption data directly,
and revise the E1 model to use the {\it vabmkl} instead of the {\it mekal} model.
The {\it vabmkl} model, an extension of the {\it mekal} model, was 
constructed for the combined fit and used the column density and plasma length as the fit parameters.
 (see \cite{yao09} for a detailed model description). 
For the halo components of the emission spectra, we fixed the abundance ratio of oxygen to
hydrogen to the solar value and allowed the abundances of nitrogen, neon,
and iron to vary again.
All parameters except for the normalization of the CXB components are linked over
the two sets of emission data. 
We put lower
and upper limits (70-440 km s$^{-1}$) to the velocity dispersion ($v_b$) which
represent the 90 \% error
range of the values obtained by the absorption analysis.

The model C1 fits  both data sets ($\chi^2$/dof=802.78/754) and 
the results are given in Table \ref{tb:FitResultsUni}.
The column density and temperature are  consistent with the A4 model (Table
\ref{tb:FitResultsAbs2}),
while the temperature and abundance of Ne and Fe are not
consistent with the E1 model (Table
\ref{tb:FitResultsEmi2}). This is because the temperature is mostly
constrained by the absorption data and the lower temperature for the
emission spectra preferred the higher abundance to describe the Ne and
Fe lines.
The plasma length is 4.2 (+1.5, $-$1.2) kpc and suggests that 
under the isothermal assumption the halo expands beyond the Galactic disk
($\sim$1 kpc).

\begin{table*}
\begin{center}  
 \caption{Combined spectral fitting results with the uniform disk model}
 \label{tb:FitResultsUni}
 \footnotesize
\begin{tabular}{llc|ccccccc|c}\hline \hline
 Model & Data & CXB & \multicolumn{7}{c}{halo} &   $\chi^2$/dof\\
& & Norm$^{a}$& $\log N_{\rm H_{Hot}}$ & $h$ & $\log T$  & 
 $v_{b}$$^{\star}$  & N/O & Ne/O & Fe/O &\\ 
& &  & (cm$^{-2}$)  & (kpc)  & (K) & (km s$^{-1}$) & &&& \\ \hline
C1 & Emission:Sz1 & $8.38^{+0.39}_{-0.38}$ &$19.08^{+0.06}_{-0.07}$ & $4.2^{+1.5}_{-1.2}$ &
			 $6.27^{+0.02}_{-0.02}$ & $\cdots$  & $4.9^{+1.4}_{-1.0}$ &
						$5.2^{+1.4}_{-1.5}$ & $5.0^{+1.6}_{-1.7}$ &802.78/754\\ 
  & Emission:Sz2 & $6.57^{+0.39}_{-0.38}$ & $\uparrow$ & $\uparrow$ & $\uparrow$ & $\cdots$
 & $\uparrow$  & $\uparrow$  & $\uparrow$ \\
 & Absorption & $\cdots$ & $\uparrow$ & $\cdots$ & $\uparrow$ & $286^{+154}_{-206}$
 & $\cdots$ & $\cdots$ & $\cdots$ &\\
\hline
\multicolumn{11}{l}{$\uparrow$ indicates linked parameters} \\
\multicolumn{11}{l}{model C1: {\it wabs}({\it power-law+{\it vmekal}})+{\it mekal} for the emission, 
{\it wabs}({\it power})$\times$({\it absem})$^{3}$ for the absorption} \\
\multicolumn{11}{l}{$^{\star}${\footnotesize Parameter range is limited to 70-440 km s$^{-1}$}} \\
\multicolumn{11}{l}{$^{a}${\footnotesize in unit of photons cm$^{-2}$ s$^{-1}$
 str$^{-1}$ eV$^{-1}$ @1keV}} \\
\end{tabular}
 \end{center}
\end{table*}

\subsubsection{Exponential Disk Model}
Observations of edge-on
galaxies(ex. \cite{wang03}, \cite{li2008}, \cite{yamasaki09}) have revealed that the intensities of
X-ray emission from extended hot gas decreases exponentially as a
function of height from the galactic plane.
As a next step in our analysis, we employed another simple model to fit the data: an  exponential distribution
model (\cite{yao09}). 
In this model, the density $n$ and
temperature $T$ of the hot gas are distributed according to the following equation, 
\begin{equation}
 n = n_0 e^{-Z/h_n \xi},   \hspace{1em} T=T_0 e^{-Z/h_T
  \xi}, \hspace{1em} \gamma=h_T/h_n
\end{equation}
where $Z$ is the vertical distance from the Galactic plane, $n_0$
and $T_0$ are the density and temperature at the plane, and $h_n$ and
$h_T$ are the scale heights of the density and temperature,  respectively,
 and $\xi$ is the filling factor, which is
assumed to be 1 in this paper. 
Thus the equivalent hydrogen column density of the hot gas ($N_{\rm H_{Hot}}$) is
calculated as $N_{\rm H_{Hot}}=\int_0^{\infty} n dl=\int_0^{\infty} n_0 {\rm
exp}(- Z/h_n)dZ/{\rm sin}b= n_0 h_n / {\rm sin}b$.

The models {\it vabmkl} and {\it absem}   
can also be used in an exponential disk model using the additional
parameter $\gamma$ (see \citet{yao09} for detailed description). 
We therefore used the same model as used in the uniform model here (model C2).
For fit parameters, for convenience we used the column density $N_{\rm H_{Hot}}$ instead of $n_0$. 

We jointly fitted the emission and absorption data using this
exponential disk model. 
The parameters obtained are summarized in Table 7. We first fixed the
velocity dispersion ($v_{b}$) at 290 km s$^{-1}$.  We next examined the
robustness of the temperature ($T_{0}$), column density ($N_{\rm
H_{HOT}}$),
 and scale height ($h_{n}$), as a function of $\gamma$, $v_{b}$,
and the intensity of foreground SWCX intensity. We found that all
parameters are consistent to within 90\% statistical errors.  When we
fitted with $v_{b}$ allowed to vary freely, the best-fit value of $v_b$ became
54$^{+19}_{-13}$ km s$^{-1}$. Though  this is above the thermal velocity
($\sim$ 30 km s$^{-1}$),  it is a smaller value than that obtained from the
absorption spectrum which determined  the ratio between the OVII
K$_{\alpha}$ and K$_{\beta}$ lines. In the exponential disk model, low
($3\times 10^{5} {\rm K} < T < 10^{6}$ K) temperature plasma can exist in the
outer regions, which  contribute only to the OVII absorption
line. This might cause the smaller $v_{b}$ value. 
The
cooling time of such low  temperature  plasmas is very short, and the actual
situation will not follow such a simple exponential model in this temperature
range.  We therefore fixed $v_{b}$ at 290 km s $^{-1}$, as the best-fit
value from the absorption analysis.

Confidence contours of $h_{n}$, $T_{0}$ and $N_{\rm H_{Hot}}$ 
versus gamma are plotted in figure 7, over-laid on those of the LMC X-3 direction
\citep{yao09}. 
We then obtained the scale height for the temperature gradient as $h_{t} = 5.6^{+7.4}_{ -4.2}$
kpc and the gas density at the galactic plane as $n_{0} = (1.4^{+0.5}_{-0.4})\times 10^{-3}$ cm$^{-3}$ 
(Figure 8). This values is typical for the mid-plane plasma density \citep{Cox2005}.
As the high temperature plasma close to the Galactic plane can emit 
Fe and Ne lines efficiently, 
the spectrum can be fitted 
 without an abundance of heavy element higher than the solar value.
The emission weighted temperature calculated with best fitted parameters
using the intensity ratio of 
O\emissiontype{VIII} to O\emissiontype{VII} becomes $2.2 (+0.1, -0.1)$
$\times 10^{6}$ K.

\begin{table*}
\begin{center}  
 \caption{Combined spectral fitting results with the exponential disk  model}
 \label{tb:FitResultsNonUni}
 \footnotesize
\begin{tabular}{llc|cccccccc|c}\hline \hline
 Model & Data & CXB & \multicolumn{8}{c}{halo} &   $\chi^2$/dof\\
& & Norm$^{a}$ & $\log N_{\rm H_{Hot}}$ & $h_n$
				 & $\log T_0$  & 
 $v_{b}$$^{\star}$  & $\gamma$ & N/O & Ne/O & Fe/O &\\
& &  & (cm$^{-2}$)  & (kpc)  & (K) & (km s$^{-1}$) & &&& \\ \hline

 C2 & Emission:Sz1 &$8.26^{+0.36}_{-0.37}$  &$19.10^{+0.08}_{-0.07}$  & $2.3^{+0.9}_{-0.8}$ & 
  $6.40^{+0.09}_{-0.05}$& $\cdots$ &  
 $2.44^{+1.11}_{-1.41}$ & $5.8^{+1.6}_{-1.3}$ &
$3.1^{+1.6}_{-1.2}$ &  $1.5^{+1.0}_{-0.7}$ &\\
 & Emission:Sz2 & $6.46^{+0.36}_{-0.36}$ & $\uparrow$  & $\uparrow$ & 
 $\uparrow$ & $\cdots$ &  
 $\uparrow$ & $\uparrow$ &  
 $\uparrow$ & $\uparrow$ &\\
  & Absorption & $\cdots$ & $\uparrow$ & $\cdots$ & 
 $\uparrow$ & 290 (fixed) &  
 $\uparrow$ & $\cdots$ &  
 $\cdots$ & $\cdots$ & 792.76/757\\ 

C2 & Emission:Sz1 & $8.20^{+0.39}_{-0.42}$  & $19.13^{+0.07}_{-0.07}$ & $2.2^{+0.5}_{-0.7}$ & 
  $6.48^{+0.04}_{-0.04}$& $\cdots$ &  
 1.0(fixed) & $6.1^{+1.8}_{-1.4}$ &
$2.4^{+0.9}_{-0.9}$ & $1.0^{+0.6}_{-0.4}$ &\\
 & Emission:Sz2 & $6.40^{+0.38}_{-0.41}$ & $\uparrow$  & $\uparrow$ & 
 $\uparrow$ & $\cdots$ &  
 $\uparrow$ & $\uparrow$ &  
 $\uparrow$ & $\uparrow$ &\\
  & Absorption & $\cdots$ & $\uparrow$ & $\cdots$ & 
 $\uparrow$ & 290 (fixed) &  
 $\uparrow$ & $\cdots$ &  
 $\cdots$ & $\cdots$ & 795.64/758\\

C2 & Emission:Sz1 & $8.25^{+0.33}_{-0.38}$  & $19.10^{+0.07}_{-0.07}$ & $2.4^{+0.9}_{-0.7}$ & 
  $6.38^{+0.02}_{-0.03}$& $\cdots$ &  
 3.5(fixed) & $5.6^{+1.1}_{-1.3}$ &
$3.3^{+1.2}_{-0.8}$ & $1.7^{+0.3}_{-0.5}$ &\\
 & Emission:Sz2 & $6.45^{+0.33}_{-0.37}$ & $\uparrow$  & $\uparrow$ & 
 $\uparrow$ & $\cdots$ &  
 $\uparrow$ & $\uparrow$ &  
 $\uparrow$ & $\uparrow$ &\\
  & Absorption & $\cdots$ & $\uparrow$ & $\cdots$ & 
 $\uparrow$ & 290 (fixed) &  
 $\uparrow$ & $\cdots$ &  
 $\cdots$ & $\cdots$ & 793.64/758\\ 
\hline
\\

 C2 & Emission:Sz1 & $8.17^{+0.37}_{-0.38}$ & $19.41^{+0.19}_{-0.16}$ & $5.1^{+3.9}_{-4.8}$
 & $6.51^{+0.16}_{-0.10}$ & $\cdots$ &  
$0.43^{+1.16}_{-0.23}$ & $5.7^{+1.5}_{-1.3}$  &
$2.3^{+1.0}_{-1.0}$ &$1.0^{+0.4}_{-0.5}$ &\\
 & Emission:Sz2 &$6.37^{+0.37}_{-0.38}$  & $\uparrow$  & $\uparrow$ & 
 $\uparrow$ & $\cdots$ &  
 $\uparrow$ & $\uparrow$ &  
 $\uparrow$ & $\uparrow$ &\\
  & Absorption & $\cdots$ & $\uparrow$ & $\cdots$ & 
 $\uparrow$ & 70 (fixed) &  
 $\uparrow$ & $\cdots$ &  
 $\cdots$ & $\cdots$ & 789.75/757\\ 
\\

 C2 & Emission:Sz1 & $8.27^{+0.38}_{-0.33}$ & $19.13^{+0.06}_{-0.07}$ & $1.9^{+0.5}_{-0.3}$
&  $6.40^{+0.04}_{-0.05}$ & $\cdots$ &  
 $2.84^{+1.34}_{-1.64}$ & $5.8^{+1.7}_{-1.3}$ &
$3.2^{+1.1}_{-1.0}$ &  $1.6^{+0.5}_{-0.8}$ &\\
 & Emission:Sz2 & $6.47^{+0.38}_{-0.32}$ & $\uparrow$  & $\uparrow$ & 
 $\uparrow$ & $\cdots$ &  
 $\uparrow$ & $\uparrow$ &  
 $\uparrow$ & $\uparrow$ &\\
  & Absorption & $\cdots$ & $\uparrow$ & $\cdots$ & 
 $\uparrow$ & 440 (fixed) &  
 $\uparrow$ & $\cdots$ &  
 $\cdots$ & $\cdots$ & 795.44/757\\ \hline
\\

 C2$^{\dagger}$ & Emission:Sz1 & $8.31^{+0.33}_{-0.41}$ &$19.08^{+0.05}_{-0.07}$ & $1.5^{+0.6}_{-0.5}$ & 
 $6.36^{+0.03}_{-0.07}$ & $\cdots$ &  
 $3.39^{+2.40}_{-1.97}$ & $4.9^{+1.1}_{-0.9}$ &  
  $3.2^{+0.7}_{-0.5}$ &  $2.0^{+0.9}_{-0.8}$ &\\
 & Emission:Sz2 & $6.52^{+0.33}_{-0.41}$ & $\uparrow$  & $\uparrow$ & 
 $\uparrow$ & $\cdots$ &  
 $\uparrow$ & $\uparrow$ &  
 $\uparrow$ & $\uparrow$ &\\
  & Absorption & $\cdots$ & $\uparrow$ & $\cdots$ & 
 $\uparrow$ & $290$(fixed) &
 $\uparrow$ & $\cdots$ &  
 $\cdots$ & $\cdots$ & 803.71/757\\\hline
\\

C2$^{\ddagger}$ & Emission:Sz1 & $8.16^{+0.38}_{-0.39}$ &$19.16^{+0.09}_{-0.08}$ & $3.7^{+1.8}_{-1.1}$ & 
 $6.51^{+0.10}_{-0.07}$ & $\cdots$  &  
 $1.47^{+0.52}_{-1.05}$ & $7.2^{+2.4}_{-1.8}$ &  
  $2.2^{+1.3}_{-1.0}$ & $0.9^{+0.6}_{-0.4}$  &\\
 & Emission:Sz2 & $6.36^{+0.38}_{-0.38}$ & $\uparrow$  & $\uparrow$ & 
 $\uparrow$ & $\cdots$ &  
 $\uparrow$ & $\uparrow$ &  
 $\uparrow$ & $\uparrow$ &\\
& Absorption & $\cdots$ & $\uparrow$ & $\cdots$ & 
 $\uparrow$ & $290$(fixed) &
 $\uparrow$ & $\cdots$ &  
 $\cdots$ & $\cdots$ & 783.37/757\\

\hline
\multicolumn{12}{l}{{\footnotesize $\uparrow$ indicates linked parameters}} \\
\multicolumn{12}{l}{{\footnotesize model C2: {\it wabs}({\it power-law+{\it vabmkl}})+{\it mekal} fot the emission, 
{\it wabs}({\it power})$\times$({\it absem})$^{3}$ for the absorption}} \\
\multicolumn{12}{l}{$^{\dagger}${\footnotesize Emission measure of {\it mekal}$_{LHB+SWCX}$ is set to 0 as the lower limit.}} \\
\multicolumn{12}{l}{$^{\ddagger}${\footnotesize Emission measure of {\it mekal}$_{LHB+SWCX}$ is set to
 upper limit which corresponds to 3.5 LU O\emissiontype{VII}
 K$_{\alpha}$  emission}} \\
\multicolumn{12}{l}{$^{a}${\footnotesize in unit of photons cm$^{-2}$
 s$^{-1}$ str$^{-1}$ eV$^{-1}$ @1keV}} \\
\end{tabular}
\end{center}
\end{table*}

\begin{figure}
\FigureFile(80mm, 150mm){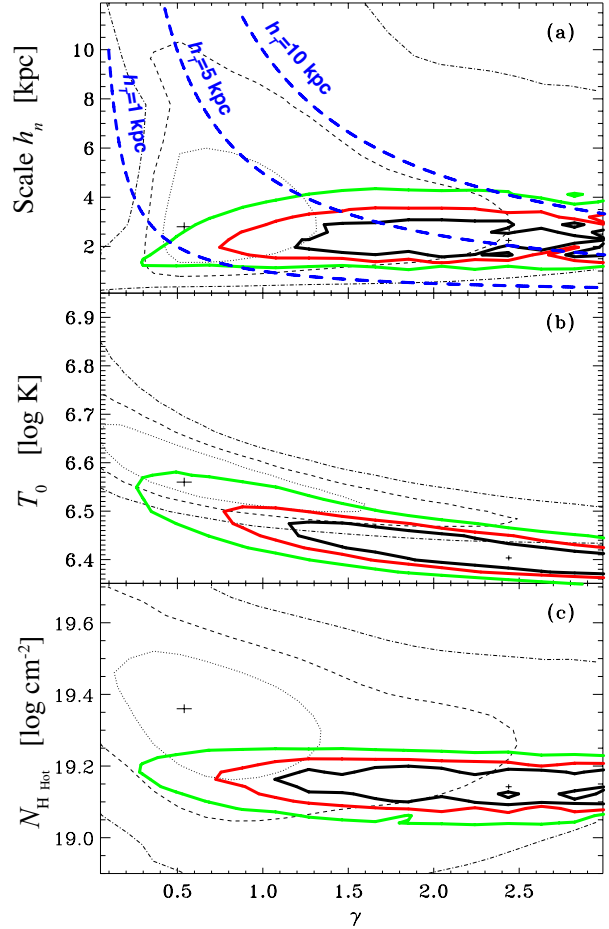}
\caption{68\%, 90\%, and 99\% confidence contours of $h_n$, $T_0$, and $N_{\rm H_{Hot}}$
 vs. $\gamma$, obtained from the combined fits
 to the X-ray absorption and emission data. Colored thick lines are for the 
 PKS 2155-304 sight line, while the black thin lines are for the  LMC
 X-3 sight lines (\cite{yao09}). In the panel (a) the scale height of
 the temperature ($h_T$) is constant along the dashed lines.}
\label{fig:Contour} 
\end{figure}

\section{Discussion}
\label{sec:discussion}

\subsection{Uncertainty due to  of  LHB and SWCX }
\label{sec:LHBandSWCX}
Because 
our knowledge about the temporal and spatial variability of the SWCX and
the LHB is limited, there are uncertainties due to the assumption of
their intensity. These uncertainties could result in large uncertainties
in our results.

To assess this uncertainty, 
we estimated the lower and upper values of the LHB and  SWCX contributions 
and evaluated the parameters of the halo components again. The lower
limit of the contribution is zero. As for
the upper limit, we adopt 3.5 LU for the  OVII emission, as
 obtained by the MBM 12 shadowing observation (\cite{smith07}). 
As the heliospheric SWCX is caused by the collision between
the Solar wind and the neutral ISM, the estimated emissivity has a peak
around the ecliptic plane (\cite{kou07} , \cite{lal04}). 
MBM 12 is located at ($\lambda, \beta$)=(47.4, 2.6) in ecliptic coordinates, while
PKS2155-304 is at ($\lambda, \beta$)=(321.2, $-$16.8). 
Thus we assume that the heliospheric SWCX contribution in the 
PKS 2155-304 direction could not be larger than that for MBM12. 

The results using these lower and upper limits are shown in Table
\ref{tb:FitResultsEmi2} and Table \ref{tb:FitResultsNonUni}.
Though the best fit values are slightly changed, they are consistent
with the previous analysis.

\begin{figure}[h!]
\FigureFile(80mm,50mm){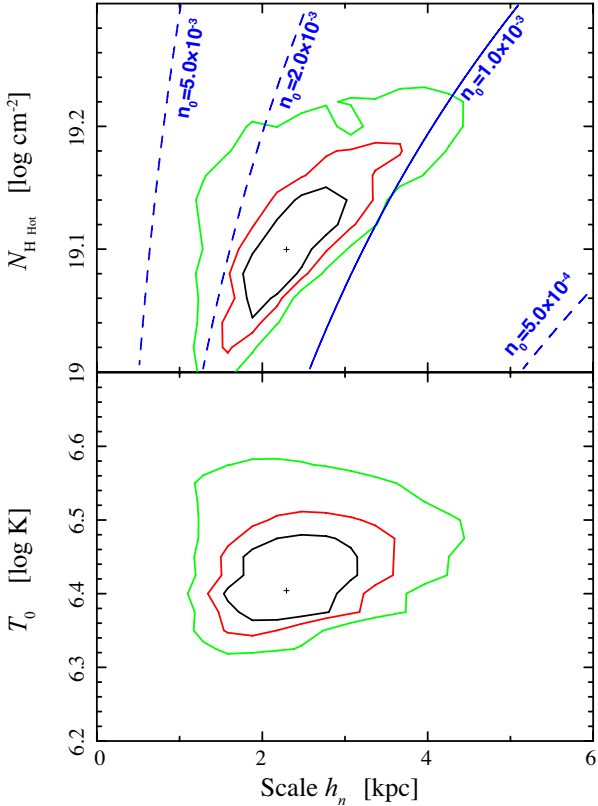}
\caption{68\%, 90\%, and 99\% confidence contours of $T_0$ and $N_{\rm H_{Hot}}$
 vs. scale height $h_n$  obtained in the joint fit
 to the X-ray absorption and emission data. 
In the upper panel the density at the plane $n_0$ is constant along the
 solid and dashed lines.
}
\label{fig:Contour2} 
\end{figure}

\subsection{Comparison with the Results for  LMC X-3}
We compared our results with those of the  LMC X-3  direction, as is
summarized in Table
\ref{tb:CompareLMCPKS}.
The directions of the LMC X-3 and PKS 2155-304 are (l,b) = (273.6,$-$32.1) and
(17.7,$-$52.2).
The fact that we obtained similar values for the two directions indicates
that the hot halo is common in the big picture and 
can be explained with the exponential model of the column density,  
scale height and temperature as $\sim 2 \times 10^{19}$ cm$^{-2}$, a few
kpc and $\sim 2 \times 10^{6}$ K.
As the distances to the targets are 50 kpc for  LMC X-3 and 480 Mpc for
PKS 2155-304, the consistency of the parameters of the exponential
disk suggests that there is little contribution from beyond LMC X-3, or from a very extended 
halo of a 100 kpc scale.

\begin{table*}[tb!]
\begin{center}  
 \caption{Disk model parameters for two sight  lines}
 \label{tb:CompareLMCPKS}
\begin{tabular}{l|ccccccc}\hline \hline
Direction & log $N_{\rm H_{Hot}}$ & $h_n$
				 & $log T_0$   & 
  $\gamma$ &Ne & Fe \\
 & (cm$^{-2}$) & (kpc) & (K)  &  & & \\\hline
PKS 2155-304 &  $19.10^{+0.08}_{-0.07}$ & $2.3^{+0.9}_{-0.8}$ & 
 $6.40^{+0.09}_{-0.05}$ &   
 $2.44^{+1.11}_{-1.41}$ &   
  $3.1^{+1.6}_{-1.2}$ & $1.5^{+1.0}_{-0.7}$  &\\
 LMC X-3$^{\dagger}$&  $19.36^{+0.22}_{-0.21}$ & $2.8^{+3.6}_{-1.8}$ & 
 $6.56^{+0.11}_{-0.10}$ & 
 $0.5^{+1.2}_{-0.4}$ &  
  $1.7^{+0.6}_{-0.4}$ & $0.9^{+0.2}_{-0.2}$  &\\
\hline
\multicolumn{7}{l}{$^{\dagger}$ {\footnotesize from \cite{yao09}}}\\
\end{tabular}
\end{center}
\end{table*}

\subsection{Distribution of the O\emissiontype{VII} and
  O\emissiontype{VIII}  Emitting/ Absorbing
  Gas and Its Origin}

\begin{figure}
\FigureFile(80mm,50mm){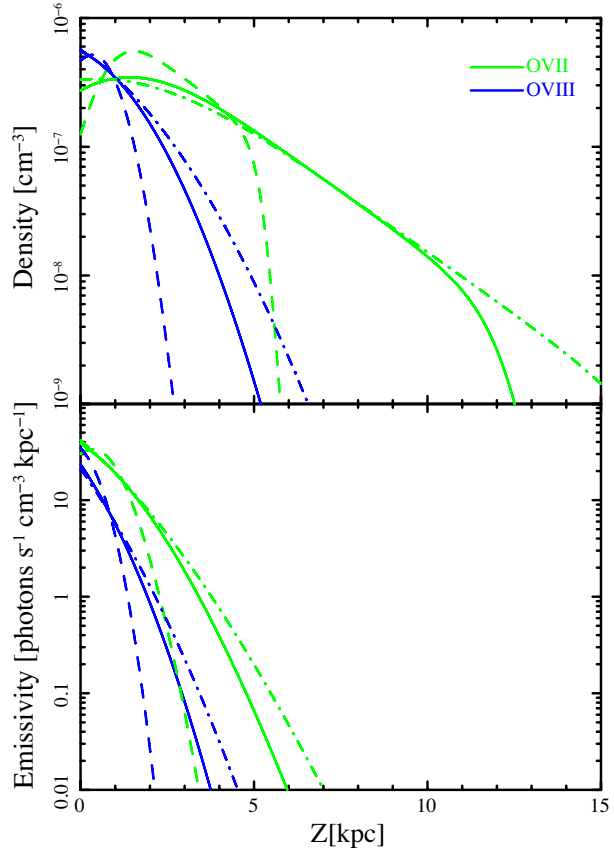}
\caption{The density of  O\emissiontype{VII} and O\emissiontype{VIII}  ion(top) and the emissivity of  O\emissiontype{VII} and
O\emissiontype{VIII}   lines (bottom) as a function of the height from the galactic
 plane under the best fit
 parameter of $\gamma$=2.44 (solid line), $\gamma$=1.0 (dashed line) and $\gamma$=3.5 (dash-dotted line).
}
\label{fig:Origins} 
\end{figure}

We calculated the distribution of  O\emissiontype{VII} and O\emissiontype{VIII}    ions and their
emissivities assuming the best fit parameters at $\gamma=2.44$ and at $\gamma= 1.0$ and 3.5 
(Fig. \ref{fig:Origins}).

\begin{figure}
\FigureFile(80mm,50mm){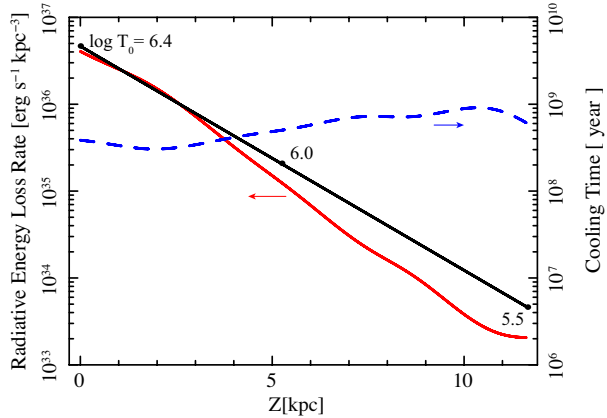}
\caption{Radiative energy loss rate (red, solid) and cooling
 time (blue, dashed) as a function of the distance  from the Galactic plane.
  The temperature is indicated by the solid black line.
 The emissivity is  calculated from the {\it mekal} ~ model,
 using a script made by Sutherland.\\
(http://proteus.pha.jhu.edu/$\sim$dks/Code/\\
Coolcurve\_create/index.html)
}
\label{fig:ELoss} 
\end{figure}

We then estimated the total radiative energy loss from the thick disk  distributed 
exponentially.
Assuming solar abundances, best fit parameters and 
ionization fraction and emissivity as taken from
SPEX\footnote{http://www.sron.nl/index.php?option=com\_content \\
\&task=view\&id=125\&Itemid=279}, 
we obtained the energy loss rate as a function of the distance  from the
Galactic plane $Z$ (Fig. \ref{fig:ELoss}).
We then integrated the energy loss rate until the temperature of the
exponential disk become lower than $10^{5.5}$ K.
Because our results are based on X-ray observations, it is difficult to detect
plasma of T $< 10^{5.5}$ K. 
We obtained a total radiative energy
loss rate of $7.2\times 10^{36}$ erg s$^{-1}$ kpc$^{-2}$ in 0.001--40 keV
and $1.8\times 10^{35}$ erg s$^{-1}$ kpc$^{-2}$ in 0.3--8.0 keV.
These values are consistent with the X-ray luminosity of other spiral galaxies \citep{strickland04}.

We next compared the energy loss rate with the energy input rate from SNe.
According to \citet{fer98}, the SN rate near the sun is
19  Myr$^{-1}$ kpc$^{-2}$ for type II SNe and 2.6 Myr$^{-1}$ kpc$^{-2}$ for type Ia SNe, respectively.
Assuming each SN explosion releases 1 $\times 10^{51}$ ergs, the total
input energy is then 7 $\times 10^{38}$ ergs s$^{-1}$ kpc$^{-2}$.
If 1 \% of the SN explosion energy is input to the hot halo, 
the total energy loss can be compensated.

\subsection{Consistency with O\emissiontype{VI} Absorbing Gas}
It is not clear that 
our model is consistent beyond $\sim 5$ kpc where the temperature of
the gas is below $\sim 10^{6.0}$ K and O\emissiontype{VI} ion becomes dominant.

\citet{wil07} found two local O\emissiontype{VI} absorption lines 
in the {\it FUSE} PKS 2155-304 spectrum 
and 
reported column densities of 1.10$\pm 0.1 \times 10^{14}$ and 
8.7$\pm 0.4 \times 10^{13}$ cm$^{-2}$.  
Our exponential disk model expects OVI
column densities of $3.8\times 10^{13}$, $1.4\times10^{14}$, and
$2.1\times10^{13}$ cm$^{-2}$ with the best fit parameters when
$\gamma$=2.44, 1.0, and 3.5 respectively.

However
 a plasma emitting O\emissiontype{VI}  lines 
cools very rapidly and 
it would be difficult to maintain such plasma existing high above the
Galactic plane.
Radiative cooling is accelerated by the density fluctuations.
Thus OVI absorbing gas can be a patchy or blob-like condensation.
To discuss this problem, energy and matter flow models are needed, which
is beyond the focus of this paper.

\section{Summary}
We have analyzed high resolution X-ray absorption/emission data observed by
{\it Chandra}  and {\it Suzaku}  to determine the physical state of the global hot gas 
along the PKS 2155-304 direction.

\begin{enumerate}
 \item Suzaku clearly detected O\emissiontype{VII} K$_{\alpha}$, O\emissiontype{VIII} K$_{\alpha}$ and 
             O\emissiontype{VII} K$_{\beta}$ lines.
	   The surface brightnesses  of O\emissiontype{VII} and O\emissiontype{VIII} in this direction can be
	   understood in the same scheme as  obtained by 
	   other 14 observations (\cite{yoshino09}).
 \item By the absorption analysis, column density is measured as 3.9
	   ($+0.6,-0.6$) cm$^{-3}$ pc
	   and temperature is measured as 1.91 ($+0.09,-0.09$) $\times 10^6$ K.
	   By the emission analysis, emission measure is measured as 3.0 ($+0.3,-0.3$) $\times 10^{-3}$
	   cm$^{-6}$ pc and temperature is measured as
	   2.14 ($+0.15,-0.14$) $\times10^{6}$ K.
 \item Combined analysis using the exponential disk model gives a good fit  
	   with $\chi^2$/dof of 789.65/756 to both
	   emission and absorption spectra. The  gas temperature and density at the
	   Galactic plane are determined to be
	   $2.5 (+0.6,-0.3)\times 10^{6}$ K and $1.4(+0.5,-0.4)\times 10^{-3}$ cm$^{-3}$
	   and the scale heights of the gas temperature and density
	   $5.6 (+7.4,-4.2)$ kpc and
	   $2.3 (+0.9,-0.8)$ kpc, respectively. 
 \item The results obtained by the combined analysis are consistent with those 
	  for  the LMC X-3 direction. This suggest that the global hot gas
	  	   surrounding our Galaxy has common structure.
\end{enumerate}

\bigskip
Part of this work was financially supported by 
Grant-in-Aid for Scientific Research (Kakenhi) by MEXT, 
No.  20340041, 20340068,  and 20840051.
TH appreciates   the support from the JSPS research fellowship and 
the Global COE Program "the Physical Sciences Frontier", MEXT, Japan



\end{document}